# Nonlinear master relation in microscopic mechanical response of semiflexible biopolymer networks

N. Honda[1], K. Shiraki[1], F. van Esterik[1], S. Inokuchi[1], H. Ebata[1], and D. Mizuno[1]

Email: mizuno@phys.kyushu-u.ac.jp

[1]Department of Physics, Kyushu University, 819-0395 Fukuoka, Japan.

**Abstract**

A network of semiflexible biopolymers, known as the cytoskeleton, and molecular motors play fundamental mechanical roles in cellular activities. The cytoskeletal response to forces generated by molecular motors is profoundly linked to physiological processes. However, owing to the highly nonlinear mechanical properties, the cytoskeletal response on the microscopic level is largely elusive. The aim of this study is to investigate the microscopic mechanical response of semiflexible biopolymer networks by conducting microrheology (MR) experiments. Micrometer-sized colloidal particles, embedded in semiflexible biopolymer networks, were forced beyond the linear regime at a variety of conditions by using feedback-controlled optical trapping. This high-bandwidth MR technology revealed an affine elastic response, which showed stiffening upon local forcing. After scaling the stiffening behaviors, with parameters describing semi-flexible networks, a collapse onto a single master curve was observed. The physics underlying the general microscopic response is presented to justify the collapse, and its potentials/implications to elucidate cell mechanics is discussed.

**Introduction**

The mechanics of living cells and tissues are governed by the networks of biopolymers that pervade in and between cells. These biopolymers differ from traditional synthetic ones in that they are semiflexible. The semifixibility results in a remarkable nonlinear mechanical response commonly shared by the biopolymers. They rapidly stiffen when they are slightly stretched, and they abruptly soften to almost zero resilience when they are compressed [1, 2]. This sudden weakening response to compression may look similar to the Euler's buckling transition which occurs for thin rods with macroscopic dimensions [3]. However, when a filament has microscopic dimensions, its nonlinear mechanical response occurs for a different reason. Besides, the crossover between weakening and stiffening of semiflexible polymer is continuous whereas buckling transition for a macroscopic rod is discontinuous. A microscopic biopolymer experiences thermal fluctuations. Stretching and compression of the filaments regulate their conformational entropy that gives rise to highly nonlinear mechanical response [4, 5]. When these filaments with nonlinear mechanical properties form a network, their macroscopic mechanical properties (shear elasticity) also exhibit nonlinearity. Remarkably, a few percentage of shear strain stiffens gels composed of semiflexible biopolymers [6-8], which is not observed in gels made of flexible polymers [7, 8].

In living cells, mechanical loading on cytoskeletal filaments is conducted by molecular motors. Motor proteins, such as myosin mini-filaments, crosslink cytoskeletal filaments and apply tensile forces to them by generating contractile forces [9, 10]. In tissues, cells bind to an extracellular network of biopolymers, *e.g.*, collagen, fibrin, and laminin [11, 12]. These cells transmit forces generated by molecular motors inside of their bodies, and apply tensile stresses to extracellular matrices. The network of semiflexible polymer stiffen when the tensile forces percolate through over the whole specimen [13]. This macroscopic stiffening has been observed for actin/myosin active gels [9, 14] and collagen gels contracted by living cells [12, 15], by using macro-rheometers.

In prior studies, conventional macro-rheometers have been used to obtain the shear viscoelastic modulus $G(\omega) \equiv (\delta\gamma/\delta\sigma)^{-1}$, by measuring the ratio of the strain response $\delta\gamma$ to the perturbative stress $\delta\sigma = \delta\sigma_0 \exp(-i\omega t)$. The static elastic modulus $G_0$ was obtained with sufficiently small $\delta\sigma_0$ and $\omega$ (angular frequency). When constant prestress $\sigma$ was applied in addition to small perturbative stress $\delta\sigma$ to semiflexible polymer gels, $G_0$ increases with $\sigma$ as [6, 16]

$$G_0(\sigma) = G_0(0) + A \times \sigma^{3/2}, \qquad (1)$$

indicating the static nonlinear response. Here, $A$ is a constant prefactor depending on the condition of each gel and the semiflexible polymer constituting the gel.

In last decades, the macroscopic mechanical properties of these biopolymer networks have been investigated [17]. Quantitative and analytical theories were proposed for the linear and nonlinear responses under the assumption that the networks homogeneously deform while maintaining the network's structure and affinity [18, 19]. The mechanical properties of the networks, with randomly oriented filaments, are then derived from the response of each polymer between neighboring crosslinks, as briefly summarized in Appendix A. According to the theory, the semiflexible polymer network stiffens under shear by following in Eq. (1) because filaments constituting the network stiffen nonlinearly under tension caused by the imposed shear [6]. Under homogeneous macroscopic shear, filaments are either elongated or compressed depending on their orientations to the shear, and stiffening of elongated filaments contributes more to the shear modulus in total. Under the assumption of the affinity, note that the linear shear viscoelastic modulus $G(\omega) = G'(\omega) + iG''(\omega)$ ($\omega$: angular frequency) of a network is inversely proportional to the stretch response of the semiflexible filament,

$$G^{-1}(\omega) \propto \alpha_l(\omega) \equiv \delta l/\delta\tau. \qquad (2)$$

Here, $\delta l$ is the alteration of the distance $l$ between crosslinks, and $\delta\tau$ is the tension applied to the filament by a sinusoidal function of time.

Even if a filament's contour length is fixed, the end-to-end distance of the filament depends on the thermal bending fluctuation. As shown in Fig. 1a, bending fluctuations are composed of different wavenumber modes. Fluctuations at each mode are stretched out in response to the applied tension and contribute to elongate the end to end distance by $\delta l$. The relaxation time of the bending fluctuation vastly spreads because it depends on the wavelength of the mode as given in Fig. 1a. By integrating the response at each mode, $G(\omega)$ was predicted to show the characteristic dependency $G(\omega) \propto (i\omega)^{3/4}$ at frequencies above $\omega_0$ which is the relaxation frequency for the mode with longest wavelength. On the other hand, elastic modulus $G'(\omega)$ exhibits an elastic plateau with a certain real constant $G_A$ below $\omega_0$. Because fluctuations with wavelengths greater than $l$ are neglected in the affine network model, each crosslink is constrained to move according to the macroscopic deformations subjected to the network (Fig. 1b) [20]. Therefore, the response of each filament to an applied tension is quasi-static for $\omega < \omega_0$.

Although the frequency bandwidth of conventional macrorheometers is limited, $G(\omega)$ of crosslinked biopolymer networks has been measured in a large range of frequencies with a technology referred to as microrheology (MR). At high frequencies ($\omega > \omega_0$), measured $G(\omega)$ of crosslinked actin gels were consistent to the behavior $G(\omega) \propto (i\omega)^{3/4}$ expected by the affine network model [9, 10, 21]. In this frequency range, the motion of a biopolymer network and solvent is strongly coupled; the gel behaves as a single incompressible continuum, which ensures the affinity of deformation [22, 23]. Therefore, it was reasonable that the affine theory is consistent to experimental results for $\omega > \omega_0$. At lower frequencies, weak frequency dependency $G(\omega) \propto (i\omega)^\beta$ with small $\beta$ (typically $0 < \beta < 0.3$) was observed instead of the elastic plateau $G(\omega) \sim G_A$ ($\omega < \omega_0$) expected by the affine theory [8, 24], indicating the presence of additional slow relaxations. Note that the fundamental assumptions made by the affine network theory (*i.e.*, the fixed network structure and affinity) are likely broken for slow relaxation responses in actual gels. Since biopolymer gels are weakly crosslinked by *e.g.*, hydrogen bonding, the network could gradually remodel via the glassy structural relaxations (Fig. 1c) [25, 26]. Even if the network structure was maintained, biopolymer gels subjected to mechanical perturbations could still relax by breaking the affinity of deformations. For a randomly-oriented and randomly-crosslinked network, an energy cost for a deformation would not be minimized under the restriction of affinity [20]. Although high-frequency perturbations initially invoke an affine deformation [27], filaments in the network further optimize their total stretch energy by introducing non-affine inhomogeneous deformations [27, 28]. Consequently, $G(\omega)$ of biopolymer gels shows slow relaxations that typically extend beyond the experimentally available frequencies [29]. Then, it was hard to precisely estimate $G_A$ from the observed $G(\omega)$.

In our prior MR study [8], we have shown that the affine elasticity $G_A$ and $\omega_0$ can be estimated from the measured $G(\omega)$ at the point where $G'(\omega)$ and $G''(\omega)$ intersect, *i.e.*, $G_A = G'(\omega_0) = G''(\omega_0)$. Here and hereafter, ′ and ″ indicate the real and imaginary part of the complex function, respectively. The intersection of $G'(\omega)$ and $G''(\omega)$ appears at the crossover between low-frequency [$G(\omega) \propto (i\omega)^\beta$, $G' > G''$] and high-frequency [$G(\omega) \propto (i\omega)^{3/4}$, $G' < G''$] region. The fundamental observation underlying this estimation procedure is that the affine response should be first established after the mechanical perturbations are applied to the network. Thereafter, non-affine deformations and other structural relaxations will follow [30, 31]. Here, $\omega_0$ is largely dependent on the parameters characterizing the network as $\omega_0 = (\kappa/\zeta_\perp)(\pi/l)^4$ where $\kappa$ is the bending modulus, *l* the crosslink distance, and $\zeta_\perp$ the transvers friction coefficient per filament length [19]. For typical specimen used in this study, $\omega_0/2\pi$ ranges from 1 Hz to 1 kHz as seen in Fig. 1a. The experimental bandwidth of commercial macro-rheometers is limited (< 10 Hz), and usually not sufficient to obtain $G_A$. The high-bandwidth MR technique is then suitable for measuring $G_A$.

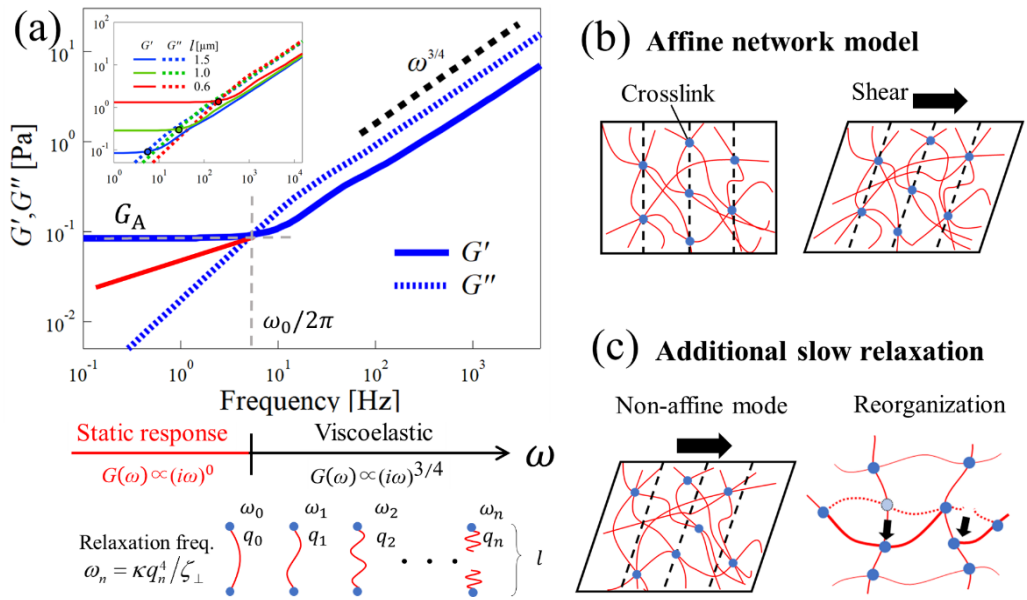

Fig. 1: (a) Theoretical prediction of the real [$G'(\omega)$: solid lines] and imaginary part [$G''(\omega)$: dotted lines] of the shear viscoelastic modulus $G(\omega)$ calculated based on the affine network model given in Ref. [19]. The following parameters consistent to the vimentin network used in this study were chosen: persistence length: $l_p = 0.8$ μm, length density: $\rho = 19 \times 10^{-12}$ m$^{-2}$, and filament diameter: $d = 10$ nm. The inset shows $G(\omega)$ calculated for different crosslink distance $l$. The circles are the intersection between $G'(\omega)$ and $G''(\omega)$ for each $l$. (b) A schematic illustration of an affine response of a semi-flexible polymer network to a simple shear deformation. The crosslink points are constrained to move in accordance with the macroscopic deformation. On the other hand, thermal bending fluctuations occur in the filaments between the crosslinks. These fluctuation can be decomposed into modes with different wave numbers $q_n = (n+1)\pi/l$ with their own relaxation frequencies $\omega_n = \kappa q_n^4 / \zeta_\perp$ ($\kappa = l_p k_B T$: bending modulus, $\zeta_\perp$: transvers friction coefficient per filament length). Consequently, $G(\omega)$ shows a power-law viscoelastic modulus $G(\omega) \propto (i\omega)^{3/4}$ for $\omega > \omega_0$, and an elastic plateau $G'(\omega) \sim G_A$ for $\omega < \omega_0$. (c) In actual gels, heterogeneous deformations that break affinity and/or reorganization of the network structures induce additional relaxations in $G(\omega)$ at $\omega < \omega_0$ as shown by the red solid line in (a).

In this study, we measured the mechanical response of biopolymer gels to forces locally applied to biopolymer networks by using the nonlinear MR technique [32]. The optical-trapping force was applied to micrometer-sized particles that were embedded in different cytoskeletons: F-actin [6], vimentin [16], and extracellular matrices of fibrin [33] reconstituted *in vitro*. A range of forces were stably applied to the probe particle embedded in the biopolymer gels, by the feedback-controlled optical trapping operated in the force-clamp mode [32]. By forcing the probe particle beyond the linear response regime, a remarkable reduction of thermal fluctuations of the probe particle was observed in the same direction as the applied force [8]. The high-bandwidth linear viscoelastic modulus $G(\omega)$ was then obtained from the thermal fluctuations by following the standard protocol for the one-particle passive MR (PMR) that relies on the fluctuation-dissipation theorem (FDT) [34]. From the high-bandwidth $G(\omega)$, the affine nonlinear elasticity $G_A(F)$ was obtained as a function of the force $F$ applied to the probe particle. $G_A(F)$ showed monotonous increase with $F$, indicating that the biopolymer gels stiffened with the locally applied force. The local stiffening occurred in a qualitatively similar manner to the stress stiffening behavior [Eq. (1)] which was measured by conventional macro-rheometers in prior studies [6]. By normalizing $G_A$ and $F$ with independently-estimated microscopic parameters ($\kappa$, *l*, $\rho$: length density and *a*: probe particle's radius), the stiffening behavior was found to collapse onto a single master curve. The distribution of stresses and strains around the locally applied force cannot be predicted with analytical theories yet because they are highly complex owing to the highly-nonlinear mechanical properties of the semiflexible polymer network. Rigorous numerical simulations were therefore necessary to investigate each experimental condition [35]. In this study, we present a qualitative theoretical explanation that verifies the existence of the single master relation, and discuss how the relation could fail at length scales smaller than those explored in this study. We conclude this article by commenting the potential of the master relation to elucidate complex mechanical properties of cells.

**Materials and Methods**

*Actin gels crosslinked with heavy meromyosin*

Globular (G)-actin was obtained from rabbit skeletal muscle as reported previously [36], and stored in G-buffer containing 2 mM tris-Cl, 0.2 mM $CaCl_2$, 0.5 mM dithiothreitol (DTT), 0.2 mM ATP, pH = 7.5 at −80ºC. Heavy meromyosin (HMM) protein was purchased from Cytoskeleton Inc., USA, and used as a crosslinker. G-actin and HMM were diluted into filamentous (F)-buffer without ATP (2 mM HEPES, 2 mM $MgCl_2$, 50 mM KCl, 1 mM EGTA, pH = 7.5) with a small amount of silica particles

(Polysciences, Inc. USA, $2a$ = 1.0 μm). The concentration of G-actin and HMM were 0.6mg/ml and 0.2mg/ml, respectively. After mixing, samples were immediately loaded on to the custom-built glass chamber made of a microscopic glass slide (size 7.6 × 2.6 cm; Matsunami Glass Ind., Ltd., Japan) and a No.1 cover slip (size 26 mm × 10 mm × 150 μm; Matsunami Glass) placed on two parallel layers of double-sided tape. Samples were left for polymerization for at least 1 h.

*Vimentin gels*

Mouse vimentin was obtained as reported previously [37-39] or purchased from Cytoskeleton Incorporated. The protein was stored in a subunit buffer containing 5 mM piperazine-N,N'-bis(2-ethanesulfonic acid) (PIPES), 1.0 mM DTT, pH = 7.0 at -80 ºC. Prior to use, vimentin was dissolved into vimentin (V)-buffer (5 mM PIPES, 1.0 mM DTT, and 270 mM NaCl) together with latex probe particles (Polysciences, $2a$ = 2.0 μm). Mixed samples were immediately infused to glass chambers and incubated on a culture rotator (RT-30 mini, TAITEC, Japan) for overnight at room temperature (R.T.). A vimentin concentration of 0.87 mg/ml was used in our study.

*Fibrin gels (fine clots)*

Human fibrinogen plasminogen-depleted protein (FIB3, Enzyme Research Laboratories, USA) was purchased and dissolved in 20 mM sodium citrate-HCl, pH = 7.4. The solution was dialyzed using a Regenerated Cellulose (RC) Membrane Spectra/Por 2 MWCO:12-14 kDa in fine clot buffer (50 mM TRIS−HCl, 400 mM NaCl, 32 mM $CaCl_2$, pH 8.5) for 2 days by exchanging the buffer twice. Then, the solution was filtered through a 0.2 μm filter (ADVANTEC, 28SP020RS). The concentration of the solution at this stage (5 ~ 8 mg/ml) was determined by measuring the UV absorption at $\lambda$ = 280 nm with a Biomate 3 spectrophotometer (Thermo Fisher Scientific, USA) using $OD_{280}$ = 1.51. A fibrin gel was prepared by polymerizing the necessary amount of fibrinogen with 0.5 U/ml bovine α-thrombin (Enzyme Research Laboratories). After quickly mixing the fibrinogen, thrombin, and a small amount of colloidal particles (silica $2a$ = 1 μm, latex $2a$ = 2 μm, Polysciences) in the fine clot buffer, the sample was immediately infused into a custom-built sealed glass chamber. The sample was incubated, while being rotated on RT-30mini (TAITEC, Japan) for preventing sedimentation of probe particles, for 2 hours at 37 °C.

The formation of a fibrin network subtly depends on buffer conditions. Under near-physiological conditions, purified fibrin forms thick bundled fibers comprised of several tens of protofibrils (coarse clots) [40]. In contrast, at high pH and ionic strength, protofibril bundling is inhibited and so-called "fine clots" are formed [41]. In order to obtain the length density $\rho$ of the fibrin network, the mass-length ratio (molecular weight per unit length of filament) of fibrin was determined by measuring the turbidity of our sample in the wavelength range of 350 nm $< \lambda <$ 900 nm [42, 43]. Experiments and analyses were conducted by following detailed procedures [33, 44, 45].

*Optical-trapping based MR*

In this section, we provide a brief overview of the optical-trapping based MR which is conducted without feedback control. Details for this conventional methodology are given elsewhere [34, 48], but repeated here to understand more advanced technique with the feedback control. As shown in Fig. 2a, a drive laser ($\lambda$ = 1064 nm, Nd:YVO$_4$, Coherent Inc., Tokyo, Japan) and a probe laser ($\lambda$ = 830 nm, IQ1C140, Power Technology, Inc., Alexander, AR, USA) were used for optical trapping of a probe particle. An acousto-optic deflector (AOD; DTSX-400-1064, AA Opto-Electronic, Orsay, France) controls the focus position of the drive laser in one direction in the sample plane. After going through the specimen, these lasers were separately detected with the quadrant photodiodes. The displacement of the probe particle from each laser focus was then obtained using the back-focal-plane laser interferometry (BFPI) technique [46, 47].

For active MR (AMR), the optical-trapping force $\delta F(t) = \delta \hat{F}(\omega)\exp(-i\omega t)$ is applied to the probe particle by the AOD-controlled drive laser, and the displacement response $\delta u(t) = \delta \hat{u}(\omega)\exp(-i\omega t)$ is measured with BFPI using the fixed probe laser. Shear viscoelastic modulus $G(\omega)$ of the medium surrounding the probe particle is then obtained from the linear response function $\alpha(\omega) \equiv \delta \hat{u}(\omega)/\delta \hat{F}(\omega)$, by using the Stokes relation extended to the frequency-dependent responses

$$G(\omega) = 1/6\pi\alpha(\omega)a. \qquad (3)$$

where $a$ is the probe particle's radius.

For passive MR (PMR), we measure the spontaneous thermal fluctuations of the probe $u(t)$ and calculate the power spectral density (PSD), $\langle |\tilde{u}(\omega)|^2 \rangle$. Imaginary part of the complex response function $\alpha(\omega) = \alpha'(\omega) + i\alpha''(\omega)$ is obtained from the PSD by using the FDT,

$$\alpha''(\omega) = \frac{\omega \langle |\tilde{u}(\omega)|^2 \rangle}{2k_B T}. \qquad (4)$$

After the real part of the response function $\alpha'(\omega)$ is obtained via the Kramers-Kronig relation [50], Eq. (3) is used to obtain $G(\omega)$. The error of $G(\omega)$ from the calculation of Kramers-Kronig relation only arise around the lower and upper limits of the measured frequency range [34].

PMR often requires a shorter period of time for each measurement compared with AMR. For PMR, one must rely on the FDT which is valid at equilibrium. In this study, the samples can be considered near equilibrium during each PMR experiment, even under constant forcing. PMR was therefore conducted in this study.

*Force feedback of the drive laser*

In this study, we attempted to conduct PMR while a constant optical-trapping force $F$ was applied to a probe particle. The optical-trapping force $F$ could not be controlled well if the focus position of the drive laser was fixed, because the probe particle fluctuated. It was also necessary to apply at maximum a ~ nN force in order to force the probe particle beyond the linear response regime. If the drive laser had been fixed, the movement of the particle would have been suppressed by the potential created by the drive laser. Therefore, the probe movement did not describe the mechanical property of the surrounding medium. In order to circumvent these problems, the drive laser was controlled with an AOD to rapidly follows the spontaneous thermal fluctuation of the probe particle. This technique, referred to as force feedback [32], was conducted in the direction of the optical-trapping force $F$, and will be explained in more detail below.

A quadrant photodiode (QPD) is placed at the back focal plane of the objective/condenser lens for the drive laser (Fig. 2a). This QPD detects the separation of the drive laser $u_d(t)$ given as $u_d(t) = C_d \times V_d(t)$, from the probe particle's center $u_p(t)$ (Fig. 2b). Here, $V_d(t)$ is the output signal from the QPD and $C_d$ is the calibration factor for the displacement response of the QPD. An electric signal $\varepsilon_{AOD}$ was generated using a 100 kHz analog proportional-integral-derivative (PID) controller (SIM960, Stanford Research Systems Inc., Sunnyvale, CA, USA) by using the integral channel as [49]

$$\varepsilon_{AOD}(t) = I \int \{V_d(t) - s(t)\} dt. \tag{5}$$

Here, $s(t) = s_0$ is the target value for the feedback control, which is set as a constant to perform the optical trapping in the force-clamp mode, and $I$ is the feedback gain of the integral channel. Since the AOD-controlled laser is moved by $u_{AOD} = C_{AOD} \times \varepsilon_{AOD}$, $u_d(t)$ can be written as

$$u_d(t) = C_{AOD} \times \varepsilon_{AOD} - u_p(t) = C_d V_d(t), \tag{6}$$

where $C_{AOD}$ is the calibration factor for the movement of the drive laser due to the deflection with the AOD. Combining Eq. (5) and (6),

$$\tilde{u}_{AOD}(\omega) = \frac{1}{1 - i\omega\tau} \tilde{u}_p(\omega), \quad \tilde{u}_d(\omega) = \frac{i\omega\tau}{1 - i\omega\tau} \tilde{u}_p(\omega), \tag{7}$$

are obtained, where ~ indicates the Fourier transformed function [$\tilde{u}(\omega) = \int_{-\infty}^{\infty} u(t) \exp(i\omega t) dt$], and $\tau \equiv C_d / IC_{AOD}$ is the cut-off frequency of the linear filtering that the feedback control performs on $u_{AOD}(t)$ and $u_d(t)$. In this study, $\tau$ was set to be ~ $10^{-5}$ s.

With this feedback control technique, the drive laser $u_{AOD}$ follows the probe particle's fluctuations in the range of $\omega < 1/\tau$. Without this linear filtering, short-time delays of apparatuses cause the resonant amplification of the signal and the feedback-response of the system will become unstable. By adjusting the feedback gain $I$, $\tau$ can be made sufficiently smaller than the characteristic response time of the probe particle in the optical-trapping potential $\tau_c \equiv \gamma / k_d$, where $\gamma \sim 6\pi\eta a$ is the friction of the probe and $k_d$ is the trap stiffness. When $\tau \ll \tau_c$, the optical-trapping force $F = k_d C_d s_0$ is stably applied to the probe particle because the drive laser follows the probe fluctuations. Therefore, this feedback-control technique is referred to as the force-clamp mode of the force feedback.

*Dual feedback PMR*

The fluctuation of the probe particle $u_p(t)$ was measured with BFPI using another QPD that detects the fixed probe laser focused on to the probe particle. Note that the QPD output is linear to the probe movement merely within ~ 200 nm from the laser focus [46]. When a constant force is applied, the probe particle makes an initial elastic displacement followed by the extremely slow drift. The drift during each PMR experiment (~30 s) is so small that the power spectral density (PSD) of the probe fluctuation is hardly affected. In order to have a statistical average of the PSD, however, the PMR experiment was repeated using the same probe particle. The probe particle could then go out of the linear range of the probe laser during the repeated measurements or immediately after the initial elastic displacement. This problem was circumvented by keeping the probe particle around the focus position of the probe laser by adjusting the 3D position of a piezoelectric stage (Nano-LP200, Mad City Labs, USA). This was done by controlling the position of the piezoelectric sample stage $u_s(t)$ with another feedback referred to as the stage feedback [49]. Briefly, using the output of a QPD that detects the fixed probe laser, an electric signal was generated using the PID-feedback controller in a manner similar to Eq. (5) with $s(t) = 0$. When the generated signal was fed to the input of the piezoelectric stage, the stage was then controlled so that the probe particle was maintained close to the laser focus.

In this study, PMR was conducted under simultaneous control of position-feedback and force-feedback by applying a constant force and for conducting BFPI with a fixed probe laser. This is referred to as the dual feedback [32]. Under dual feedback control, the total probe movement $u(t)$ in the specimen was obtained as $u(t) = u_p(t) + u_s(t)$. We applied the FDT to the power spectral density (PSD) of $u(t)$, and obtained the imaginary part of the complex response function by using Eq. (4). The shear viscoelastic modulus $G(\omega)$ was then obtained by using Eq. (3).

*Experimental test of the force-feedback MR*

In Fig. 2 (c), we compared $\alpha''(\omega)$, measured with the force-feedback AMR (blue circles), to $\omega \langle |\tilde{u}(\omega)|^2 \rangle^{\text{FF}} / 2k_B T$, measured with the force-feedback PMR (blue curve). Here and hereafter, the superscript "FF" indicates the value measured under force-feedback control. The optical-trapping force was clamped to 0 pN during all experiments. The complete agreement of AMR and PMR indicated that the FDT holds. The red circles and the red curve indicate the corresponding AMR and PMR results, measured without conducting the force feedback (*i.e.* conventional AMR and PMR). Here, the response function and the fluctuations of the probe particle, that is under the optical-trapping potential, are defined as $A(\omega) = A'(\omega) + iA''(\omega)$ and $\langle |\tilde{u}(\omega)|^2 \rangle^{\text{trap}}$, respectively. It is confirmed that the FDT holds as $A''(\omega) = \omega \langle |\tilde{u}(\omega)|^2 \rangle^{\text{trap}} / 2k_B T$. It is also seen that both the response and fluctuations are remarkably suppressed at low frequencies, owing to the optical trapping of the drive laser ($k_d = 1.1 \times 10^{-5}$ N/m). In conventional MR, the artifact derived from the optical-trapping potential is corrected as [48],

$$\alpha(\omega) = A(\omega) / [1 - k_d A(\omega)]. \tag{8}$$

when the trap potential is not too strong. The corrected result is denoted as $\omega \langle |\tilde{u}(\omega)|^2 \rangle^{\text{cor}} / 2k_B T$ given as the dark yellow curve in Fig. 2c. The corrected result agreed with the results measured by the

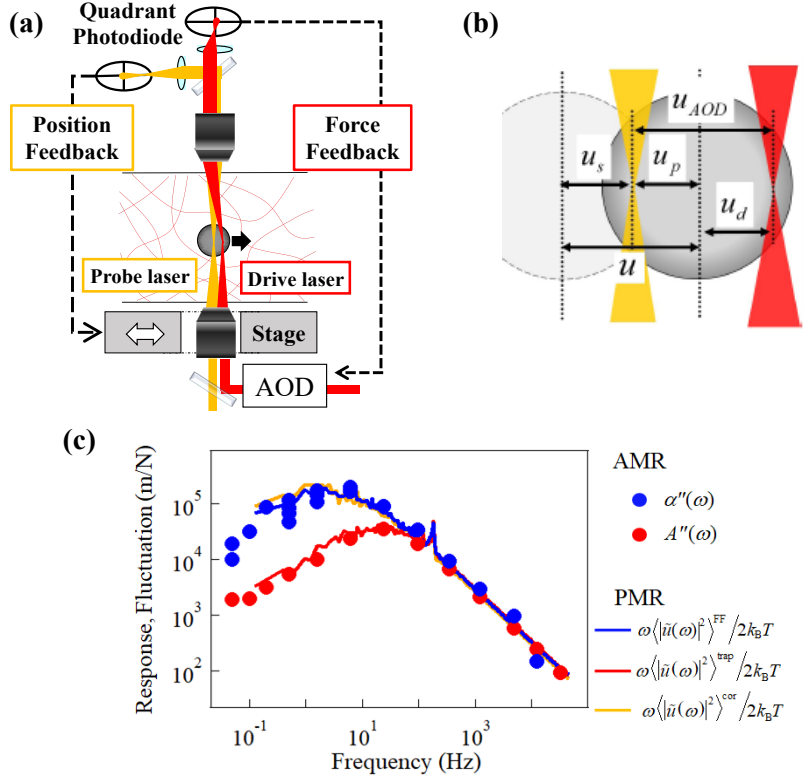

Fig. 2: (a) Experimental setup for dual-feedback MR, and (b) schematic illustration of various displacements of a probe particle. The separation between the laser focus and the center of the probe particle [$u_d$ in (b)] was measured using the QPD placed at the back focal plane of the objective/condenser lenses. The focus position ($u_{AOD}$) of the drive laser (red) was rapidly altered by the AOD to achieve the desired $u_d$ by force feedback. The weak probe laser (orange) was fixed and detected the probe particle's position ($u_p$) with another QPD. In order to maintain the probe particle within the linear range for the QPD detection, the piezoelectric sample stage was also controlled by stage feedback. (c) Result of the responses ($\alpha''$, $A''$ : circles) measured with AMR and the fluctuations ($\omega\langle|\tilde{u}(\omega)|^2\rangle/2k_BT$ : curves) measured with PMR. A $2a = 2\mu m$ latex/silica particle embedded in F-actin gel was used as a probe particle. AMR and PMR agreed for the force-feedback MR (blue symbols) and conventional MR (red symbols), respectively. The probe particle movement, measured by the conventional MRs, was suppressed at low frequencies compared with those measured with force-feedback MR. The dark yellow curve $\omega\langle|\tilde{u}(\omega)|^2\rangle^{corr}/2k_BT$ was obtained by correcting the conventional PMR data ($\omega\langle|\tilde{u}(\omega)|^2\rangle^{corr}/2k_BT$) by using Eq. (8). The consistency between conventional PMR ($\omega\langle|\tilde{u}(\omega)|^2\rangle^{corr}/2k_BT$) and the force-feedback MR ($\alpha''$, $\omega\langle|\tilde{u}(\omega)|^2\rangle^{FF}/2k_BT$) indicates that the optical-trapping potential was effectively removed by the force-feedback MR technique.

feedback AMR and PMR, confirming that the force feedback effectively removed the optical-trapping potential.

When the optical-trapping potential of the drive laser is strong and the force feedback is not performed, the complex response function $A(\omega)$ depends more on $k_d$ rather than $G(\omega)$. The artifact owing to the trapping potential prevent accurately evaluate $G(\omega)$ at low frequencies. Especially in this study, $k_d$ has to be set large compared to the ordinary MR experiments in order to force the probe particle beyond the linear response regime. It is then compulsory to effectively remove the trapping potential by conducting fast force feedback at the condition of $\tau \ll \tau_c$. Since the stage feedback is slow, it cannot completely remove the trap potential of the probe laser [49]. However, the impact of the probe laser on the probe particle motion is marginal and can be corrected by using Eq. (8) (but by exchanging $k_d$ into $k_p$) since the probe laser runs with smaller power [49, 51].

**Results**

Fig. 3 a and b show the applied-force ($F$) dependence of $\alpha''(\omega) = \omega \langle |\tilde{u}(\omega)|^2 \rangle^{\mathrm{FF}} / 2 k_\mathrm{B} T$ in vimentin and actin cytoskeletons, respectively. Probe fluctuations in the same direction to the applied force were investigated. The strong drive laser was not controlled by the force feedback in the orthogonal direction. Therefore, fluctuations in that direction were suppressed by the optical-trapping potential created by the strong drive laser. It was found that the thermal fluctuations decrease as $F$ increases. Force feedback was conducted in the same direction as that of the applied force, and we analyzed the probe particle fluctuations in that direction. When strong forces were applied with the force-clamp mode, the probe particle may gradually drift owing to the creep/plastic response of hydrogels. If such drift movement of the probe particle existed in non-negligible amount, frequency dependence typical for super diffusion, $\omega \langle |\tilde{u}(\omega)|^2 \rangle^{\mathrm{FF}} / 2 k_\mathrm{B} T \propto \omega^{-\beta}$ with $\beta > 1$, should have appeared at low frequencies and break the FDT. However, there were no such indication in the PMR spectrum in Fig. 3a and b, meaning that the drift of the probe particle was negligible in the time scale of data collection (~ 30 s) in this study. Therefore, we used the FDT to obtain $\alpha''(\omega)$ from the observed fluctuation $u(t)$ following Eq. (4). After getting the full complex response function $\alpha(\omega) = \alpha'(\omega) + i\alpha''(\omega)$ using the Kramers-Kronig relation [34, 48], the shear viscoelastic modulus $G(\omega)$ was obtained from the generalized Stokes' formula [Eq. (3)]. Eq. (3) is valid for a spherical probe particle dispersed in an "isotropic" medium whereas the network becomes mechanically anisotropic when forced in one direction [8, 27]. It may therefore be appropriate to use $\alpha(\omega)$ instead of $G(\omega)$ in the following discussion. However, for a better readership, we use the "effective" $G(\omega)$ defined by Eq. (3). In Figs. 3 (c) and (d), we show the $G(\omega)$ of a vimentin gel as a function of frequency at the applied force (c) $F$ = 0 pN and (d) $F$ = 24 pN.

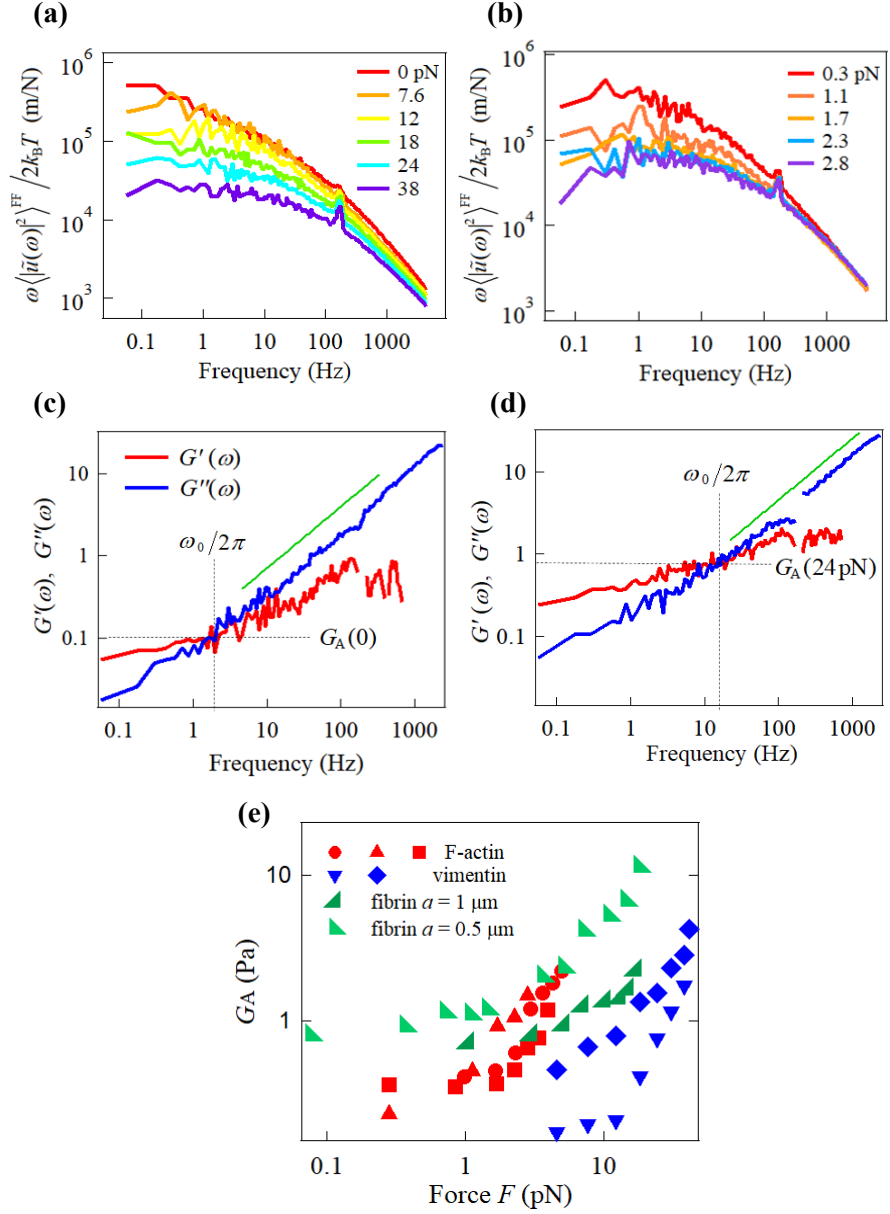

Fig. 3: PSDs of the thermal fluctuations for (a) $2a = 2$ μm latex probe particle in vimentin (0.87 mg/mL) and (b) $2a = 1$ μm silica probe particle in actin (F-actin 0.6 mg/mL crosslinked with 0.2 mg/mL HMM). Probe movements were measured at 10 kHz sampling rate with dual-feedback PMR while applying constant forces. In (a) and (b), PSDs are shown normalized as $\alpha''(\omega) = \omega \langle |\tilde{u}(\omega)|^2 \rangle^{\mathrm{FF}} / 2k_\mathrm{B}T$. Shear viscoelastic modulus of vimentin gel at $F = 0$ pN and $F = 24$ pN are shown for (c) and (d), respectively. The vertical and horizontal broken lines are drawn at the frequency $\omega_0/2\pi$ and the affine elasticity $G_\mathrm{A}(F)$, respectively, at the intersection of $G'$ and $G''$. $G_\mathrm{A}$ increased when the localized force was applied to the probe particle. Green solid lines indicate the frequency dependence ($\propto \omega^{3/4}$) expected for $\omega > \omega_0$. $G'$ (> 500 Hz) decreased due to the calculation of Kramers-Kronig relation [34]. (e) Stiffening of biopolymer networks (red: actin, green: fibrin, blue: vimentin) subjected to localized forces. The experimental conditions for vimentin and F-actin are the same as those in Fig. 3 (a) and (b). Data plotted with the same symbol indicate the set of experiments conducted using the same probe particle. Affine elasticity $G_\mathrm{A}$ scatters between different gels but shows tendency to increase with the applied force.

The affine network theory for permanently-crosslinked semiflexible biopolymers predicts that an elastic plateau appears at low frequencies as shown in Fig. 1a. However, in our experimental data, $G(\omega)$, at low frequencies, does not reveal a complete elastic plateau, but a slow viscoelastic relaxation arises as $G(\omega) \propto (i\omega)^\beta$ with $0 < \beta < 0.3$ as shown in Fig. 3c and d. These slow decays are in many cases attributed to the non-affine glassy relaxations or the other structural relaxations, such as the change in the network's topology [24-26, 30]. Whatever the mechanism of the slow relaxations is, it is reasonable to expect that these slow nonaffine modes occur after affine deformations are established in response to the external perturbation. Then, the affine response should occur [8, 27] above the crossover frequency $\omega_0$ at which the real ($G'$) and imaginary ($G''$) part of the shear viscoelastic modulus intersect. The affine elasticity, which we define as $G_A(F) \equiv G'(\omega_0) = G''(\omega_0)$ corresponds to the quantity of the elastic plateau that semiflexible biopolymer networks should have exhibited in the case of ideally affine response. As we see in Fig. 3c and d, the affine elasticity depends on the applied force $F$. The affine elasticity $G_A(F)$ was then measured as a function of the force $F$ in actin, vimentin, and fibrin gels, as shown in in Fig. 3e. All data showed stiffening upon the local force application.

To investigate the observed local stiffening $G_A(F)$, the semiflexible biopolymer networks were characterized by measuring their physical parameters as follows. The persistent length $l_p$ of F-actin (w/o phalloidin labelling) and vimentin are known to be 10 μm [52-54] and 0.8 μm [8], respectively. Length densities of the filaments $\rho$ of 0.60 mg/ml F-actin and 0.87 mg/ml vimentin are $2.4\times10^{13}$ /m$^2$ and $19\times10^{12}$ /m$^2$, respectively. The theory for the semi-flexible polymer network predicts the affine elasticity as [19]

$$G_A(0) = 6\rho k_B T l_p^2 / l^3 . \qquad (9)$$

The crosslink distances $l$ for vimentin and F-actin were then obtained from the measured $G_A(0)$ using Eq. (9). On the other hand, the diameter $d$ and the bending modulus $\kappa = l_p k_B T$ of the fibrin filament were not known because fibrin protofilaments bundle to fibers subtly depending on the solvent condition [55]. Therefore, we determined the mass-length ratio of the fibrin fiber in our sample as $\mu = 3.8\times10^{11}$ Da/cm by the turbidity measurement shown in Fig. 4a [33, 44, 45]. Then, the length density of the fibrin fiber was estimated as $\rho = 7.9\times10^{12}$ m$^{-2}$. The viscoelastic modulus of the semiflexible biopolymer network shows frequency dependency in the form of [19]

$$G(\omega) - i\omega\eta \approx \frac{1}{15}\rho\kappa l_p (i\omega/\omega_0)^{\frac{3}{4}}, \qquad (10)$$

at frequencies above $\omega_0 \equiv \kappa/2\zeta_\perp$. $\eta$ is the viscosity of the solvent. The transverse friction coefficient per unit length of the filament ($\zeta_\perp$) can be estimated from Eq. (16) in Appendix A by using the diameter $d$ of the fibrin filaments. Although the value of $d$ distributes within 15 nm $\leq d \leq$ 35 nm [33, 44], $\zeta_\perp$ weakly depends on $d$ with logarithmic function. In Fig. 4b, we show $G(\omega) - i\omega\eta$ of 0.50 mg/mL fibrin gel measured at $F = 0$. The green broken line indicates the power-law dependency ($\propto \omega^{3/4}$) that conformed to our experimental data. The fit of Eq. (10) provides the bending modulus $\kappa = l_p k_B T$ of the fibrin fiber with $l_p = 2.3 \times 10^{-6}$ m. Incorporating these values into Eq. (9), we obtain $l = 1.1 \times 10^{-6}$ m for the fibrin gels we prepared.

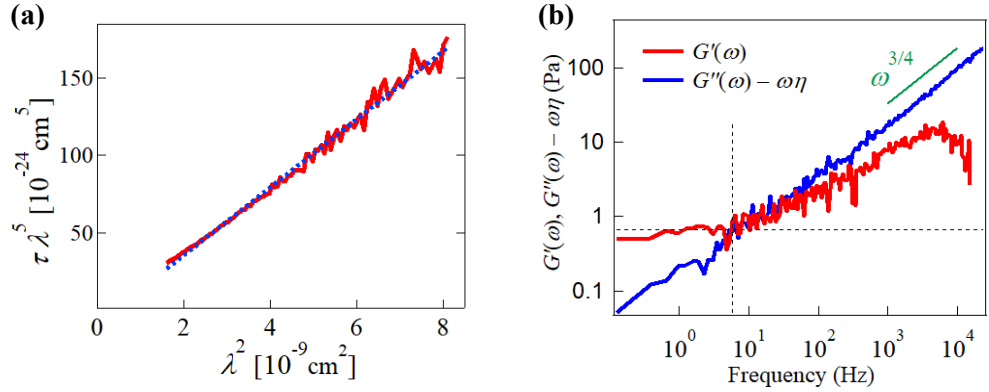

Fig. 4 (a) Turbidity $\tau \equiv \text{O.D.} \times \ln 10$ of 5.0 mg/mL fibrin polymerized in the fine-clots buffer. Except for the short wavelength region, where the light absorption rather than scattering became prominent, the linear relation between $\tau\lambda^5$ and $\lambda^2$ was observed (as expected). From the slope and the intercept at the horizontal axis, we obtained the mass length ratio $\mu = 3.8 \times 10^{11}$ Da/cm. (b) Shear viscoelastic modulus of fibrin gels (0.5 mg/mL) measured at $F = 0$ using $2a = 2$ μm latex/silica colloidal particle. Data were collected at 100 kHz sampling rate. The imaginary part is shown after subtracting the contribution from the solvent viscosity $\eta$ as $G'' - \omega\eta$. At high frequencies, we observed (as expected) a power-law behavior $G'' - \omega\eta \propto \omega^{3/4}$, as depicted by the green solid line. The decline of $G'$ at high frequency (> 5 kHz) is due to the calculation of Kramers-Kronig relation [34]. Persistent length of the fibrin fibers was obtained by fitting Eq. (10) to this result.

Using the obtained microscopic parameters, we tried to scale the measured local stiffening $G_A(F)$. In Fig. 5a, the abscissa was normalized to the dimensionless force $F^* \equiv l^2 F / \rho\kappa a^2$, and for the ordinate, $G_A(F)$ was normalized by the value $G_A(0)$ as $G_A^* \equiv G_A(F)/G_A(0)$. Although there are other choices for normalizing our data, we found that the data tended to collapse to a single master curve (Fig. 5a) with this normalization. This master relation was obtained from the experimental data with a heuristic approach. Owing to the complexity of the nonlinear response of the semiflexible polymer network, the quantitative analytical theory describing the relation is not derived yet. However, the existence of the master relation will be justified based on the theory of affine network model, in the discussion section.

The collapsed data seems to have a tendency expressed by

$$G_A^*(F^*) = 1 + (F^*/F_0)^{3/2} \tag{11}$$

with $F_0 = 14.2$. This tendency is apparently similar to the stiffening behavior measured by macro-rheometry [6], *i.e.* Eq. (1). In many macrorheometers, mechanical perturbations (stress and strain) are applied homogeneously in the sample (Fig. 5b). Eq. (11) is then equivalent to Eq. (1) for macrorheometry, as long as the same geometrical condition is employed. As shown in Fig. 5b, in the case of a parallel-plate macrorheometer, the externally applied force is given as $F = \sigma S$ where $S$ is the boundary area of sample chamber. On the other hand, in the case of MR conducted in this study, the external force was applied to a probe particle embedded in the biopolymer network. Stress and strain are heterogeneously distributed in the medium surrounding the probe particle, as shown in Fig. 5c. Semiflexible filaments in the hydrogels are subjected to either tensile or compressive stress, and then stiffen or weaken accordingly, depending on their orientation compared to the strain at each location [8, 12]. Qualitatively, the stiffening of the stretched filaments at the back of the pulled probe particle dominates over the softening of the compressed filaments at the front, leading to the overall stiffening response observed $G_A(F) \geq G_A(0)$. Nonetheless, the reason why the local stiffening also seems to follow Eq. (11), is left to be explained.

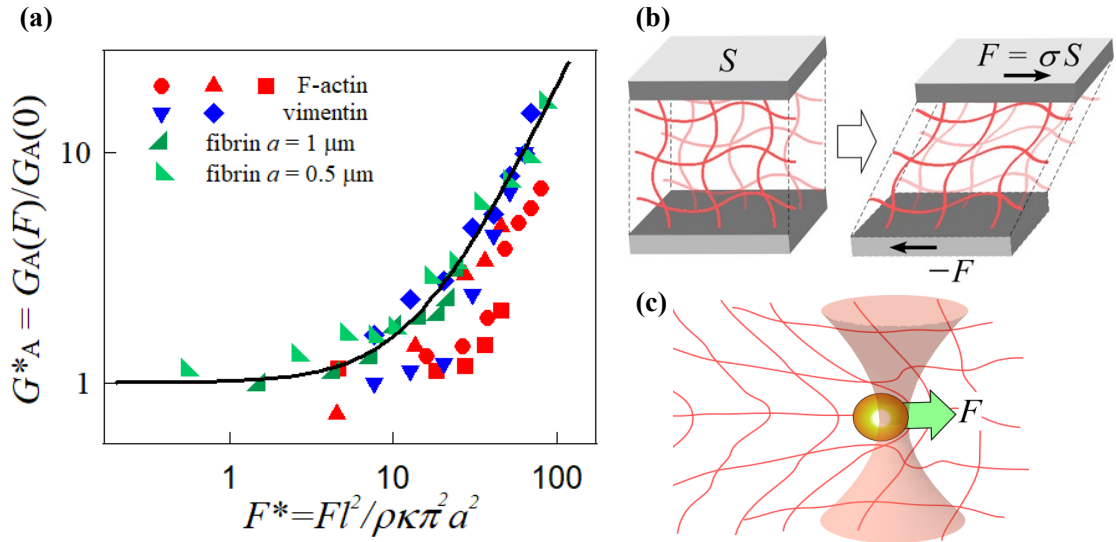

Fig. 5: (a) Affine elasticities of biopolymer gels collapsed to a single master curve when $G_A$ and $F$ are normalized to $G_A^* \equiv G_A(F)/G_A(0)$ and $F^* \equiv Fl^2/\rho\kappa\pi^2 a^2$, respectively. The same data shown in Fig. 3 (e) are plotted with the same combination of symbols. Solid curve is the fit of Eq. (11) to data for fibrin and vimentin. Actin data was excluded from the fit since they systematically deviate from others likely because of the bending energy ($\varepsilon_b$) contribution to elasticity which does not increase under force. (b) Schematic for homogeneous shear deformation in macrorheometer. (c) The network deformation by the external force $F$ applied to the probe particle is non-uniform and will be highly complex. For better visibility of the network deformation, the meshwork is sparser than the actual gel.

**Discussion**

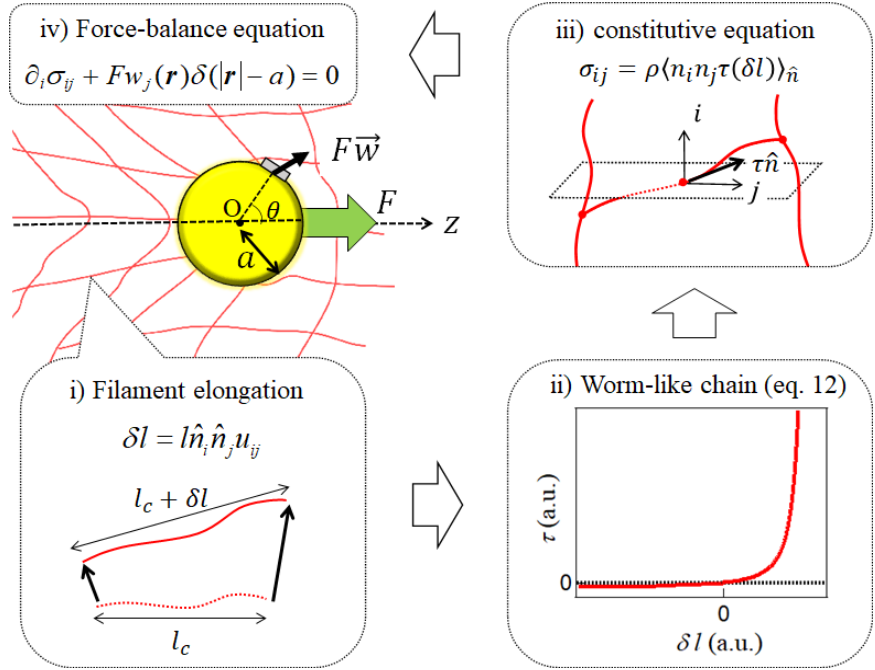

Fig. 6: The flow of theoretical analysis of the nonlinear mechanical response of semiflexible biopolymer networks. Deformation (or strain $u_{ij}$) around the optically forced probe particle depends on the position $\boldsymbol{r}$ as $u_{ij}(\boldsymbol{r})$. i) Once the filament elongation $\delta l$ is given under affine approximation, ii) tension $\tau$ applied to the filament is obtained via the worm-like chain model. iii) By taking the sum of the tension for the filaments (with orientation $\hat{n}$) that penetrate through a plane orienting at $i$ direction, a coarse-grained constitutive equation is obtained. iv) Then, a force balance equation is given by considering the force density at the boundary of the probe particle.

As reported above, thermal fluctuations of probe particles embedded in semiflexible biopolymer networks were remarkably decreased under the application of optical-trapping force, indicating stiffening of the surrounding medium. Performing experiments under various conditions of the network ($a$, $\rho$, $\kappa$, and $l$), we found that the stiffening behavior of the normalized elasticity $G_A^* \equiv G_A(F)/G_A(0)$ (affine elasticity normalized with its zero-force value) is uniquely determined by the scaled dimensionless force $F^* \equiv l^2 F/\rho\kappa a^2$. Now, we will discuss the physics underlying the collapse of the data, and justify the presence of the master relation.

Let us start with the force-extension relation of worm-like chains (WLC) given by [4]

$$\frac{\delta l}{l} = \frac{l}{6l_p} - \frac{l}{2l_p \tau^*}\left(\sqrt{\tau^*}\coth\sqrt{\tau^*} - 1\right), \tag{12}$$

where $\tau^* \equiv \tau l^2/\kappa$ is a normalized tension applied to the WLC, and $l$ is the length of the semiflexible biopolymer. For the crosslinked biopolymer network, $l$ indicates the crosslink distance. The filaments between crosslinks extend by $\delta l = l \hat{n}_i \hat{n}_j u_{ij}$ under strain $u_{ij}$ where $\hat{n}_i$ is a unit vector describing the orientation of the filament. Here and hereafter, $\hat{\ }$ indicates the unit vector. It is reasonable to introduce a normalized strain as $\hat{n}_i \hat{n}_j u_{ij}^* \equiv \hat{n}_i \hat{n}_j u_{ij} l_p/l = 1/6 - (\sqrt{\tau^*}\coth\sqrt{\tau^*} - 1)/2\tau^*$. Under the affine response, the constitutive equation for the gel made of WLC is given by $\sigma_{ij} = \rho \langle \hat{n}_i \hat{n}_j \tau(\hat{n}_k \hat{n}_l u_{kl}) \rangle_{\hat{n}}$, which is normalized to

$$\sigma_{ij}^* = \langle \hat{n}_i \hat{n}_j \tau^*(\hat{n}_k \hat{n}_l u_{kl}^*) \rangle_{\hat{n}} \tag{13}$$

with $\sigma_{ij}^* \equiv l^2 \sigma_{ij}/\rho\kappa$.

The optical-trapping force is applied to a hard sphere embedded in a medium which deforms by following the normalized constitutive equation [Eq. (13)]. The origin of the coordinate is chosen at the center of the probe particle with radius $a$, and we assume that the force is applied in $z$ direction. The external force is applied to the hydrogel medium via the boundary of the spherical probe particle. Therefore, by determining $F\boldsymbol{w}\delta(|\boldsymbol{r}|-a)$ as the force density at the probe particle surface, we obtain the force balance equation in the region of $r \geq a$,

$$\partial_i \sigma_{ij}^* + \frac{l^2}{\rho\kappa} F w_j(\boldsymbol{r})\delta(|\boldsymbol{r}|-a) = 0. \tag{14}$$

Here, $w_j(\boldsymbol{r})\delta(|\boldsymbol{r}|-a)$ denotes the polar distribution of the force at $|\boldsymbol{r}|=a$. Because of the uniaxial symmetry of the system along $z$ direction, $w_j(\boldsymbol{r})\delta(|\boldsymbol{r}|-a)$ depends solely on $\theta$ in spherical coordinate $(r,\theta,\phi)$. Since the network is assumed as continuum, the characteristic length in our experimental situation is the probe particle size $a$. Normalizing the spatial coordinate as $\boldsymbol{r}^* \equiv \boldsymbol{r}/a$, we have $\delta(|\boldsymbol{r}|-a) = \delta(|\boldsymbol{r}^*|-1)/a$, and $\partial_i \sigma_{ij}^* = \partial_i^* \sigma_{ij}^*/a$ where $\partial^*$ denotes a partial differentiation with $\boldsymbol{r}^*$. In the static situation we consider, the force-balance condition should be met at the closed surface including the probe particle as, $2\pi \int w_j \delta(r-a)r^2 \sin\theta dr d\theta = 2\pi \int w_j a^2 \sin\theta d\theta = \delta_{j,z}$. It is then appropriate to normalize $w_j(\boldsymbol{r})$ as $w_j^*(\boldsymbol{r}^*) \equiv w_j(\boldsymbol{r})a^2$. Eq. (14) is then converted to the dimensionless force-balance equation as,

$$\partial_i^* \sigma_{ij}^* + F^* w_j^*(\boldsymbol{r}^*)\delta(|\boldsymbol{r}^*|-1) = 0. \tag{15}$$

by using the normalized external force $F^* \equiv l^2 F/\rho\kappa a^2$.

The flow of the theoretical analysis procedure, as explained above, is schematically shown in Fig. 6. A combination of Eqs. (13) and (15) fully addresses the mechanical response of the surrounding medium of the probe particle to the external force $\boldsymbol{F}$. For instance, one can obtain the stress-strain distribution around the probe particle to which the force is applied with numerical simulations [35]. Note that there is only one dimensionless parameter in the set of Eqs. (13) and (15), i.e., $F^*$. Once $F^*$ is given, the polar distribution $w_j^*(\boldsymbol{r}^*)$ is also determined to satisfy the no-slip boundary condition at the probe particle's surface. This means that the scaled mechanical response does not depend on parameters, such as $\kappa, l, \rho$ and $a$, regardless of the complexity of nonlinear response of the system to the external force $\boldsymbol{F}$. In this study, we probed the mechanical response of the system by measuring the displacement $u(t)$ of the embedded probe particle. By normalizing the displacement as $u^* \equiv u/a$, the normalized stiffening response is written as $G_A(F)/G(0) = \left(\delta F^*/\delta u^*|_{F^*}\right)/\left(\delta F^*/\delta u^*|_{F^*=0}\right)$. It is now understood that this quantity is determined solely by the dimensionless force $F^*$ because $\delta F^*/\delta u^*|_{F^*=0}$ is a mere constant and $\delta F^*/\delta u^*|_{F^*}$ depends only on $F^*$.

MR performed in this study investigated the fluctuation of a single probe particle, referred to as 1-particle MR. One-particle MR provides results consistent to macro-rheometry when a specimen is homogeneous including the region close to the probe particle's surface. This condition has been usually met for various complex fluids, such as microemulsions [56, 57], glasses [58, 59], hydrogels made of flexible polymers [60-62], etc. However, inhomogeneity such as the aggregation or depletion of solutes at the probe particle surface influences 1-particle MR [63, 64]. Except such obvious effects, it has been reported that 1-particle MR conducted in the semiflexible polymer networks sometimes disagrees from macro-rheometry. Examples include non-crosslinked actin [63, 65], microtubule solutions [66], etc. For such samples, MR that investigates the correlation between 2 separate probe particles (2-particle MR) provides results consistent to macro-rheometry because 2-particle MR probes the mechanical response of a larger volume between 2 probe particles.

The inconsistency between 1-particle MR and macro-rheometry in some of semiflexible polymer networks is not trivial and may be caused by the strain gradient. The strain gradient appears in the higher order term in the expansion of strains. In the case of MR, the strain gradient naturally arises in the medium close to the embedded probe particle when the particle is forced and moved. Even in the case of PMR, the local strain gradient appears due to the thermal fluctuation of the probe particle. Filaments in the network close to the probe particle must deform (bend) according to the field. Note that the analysis, summarized in Fig. 6, is based on the traditional strain-based mechanics theory, which does not account for contributions from strain gradients. However, such local bending deformation requires extra energy compared to the estimated energy based on the affine response. Let us evaluate the local bending energy caused by a displacement $u_0$ of a probe particle from its equilibrium position. The displacement field in the surrounding medium of the probe particle scales with distance $r$ from the probe center as $\sim a u_0 / r$. The strain and strain gradient (magnitude of local bending) of the medium are then estimated as $\sim a u_0/r^2$ and $\sim a u_0/r^3$, respectively. The corresponding energy densities scale as $\varepsilon_s \sim G_A \left(a u_0/r^2\right)^2$ for the affine shear deformation and $\varepsilon_b \sim \rho \kappa \left(a u_0/r^3\right)^2$ for the local bending deformation. Around the probe particle ($r \sim a$), $\varepsilon_b$ overcomes $\varepsilon_s$ when $a \ll \sqrt{l^3/6 l_p}$ is satisfied. Thus, when smaller probe particles are used [65], the energy cost $\varepsilon_b$ becomes non-negligible, meaning that $\varepsilon_b$ is required in addition to $\varepsilon_s$ for the probe particle to move. In that case, 1-particle MR overestimates the elasticity of the network.

The local bending effect for the non-crosslinked gels was typically observed because $G_A$ and $\varepsilon_s$ are decreased owing to the lack of crosslinking. For instance, for non-crosslinked actin gels reported in Ref. [65], we estimate $\varepsilon_b \gg \varepsilon_s$ at $r = a$, leading to the non-trivial inconsistency between 1-particle and 2-particle MR. On the other hand, all of the gels used in this study are cross-linked either artificially (actin) or spontaneously (vimentin and fibrin). For these specimens, $\varepsilon_b \ll \varepsilon_s$ was satisfied at $r \sim a$ for vimentin and fibrin gels, and $\varepsilon_b \sim \varepsilon_s$ for the crosslinked actin gel. Note that $\varepsilon_b$ decays faster than $\varepsilon_s$ with the distance from the probe particle. Therefore, our samples may not be remarkably affected by the local bending deformation of the filaments. Consistently, we did not see clear-cut disagreements between 1-particle and 2-particle MR for crosslinked specimens [10, 67]. Nevertheless, the local bending energy might slightly contribute to the elasticity of our actin gels, resulting in the overestimation of $G_A(0)$. The contribution to the elasticity derived from the local bending energy will not increase under the application of the external force $F$. Then, the normalized elasticity of the actin gel data reasonably deviates from other data, distributing below the master curve as seen in Fig. 5a.

In our prior study [8], we observed stiffening of vimentin gels by placing another probe particle at the location (*r*) which was distant from where the optical-trapping force was applied. In such situation ($r \gg a$), the force applied to the network can be regarded as a monopole. Although the distribution of the force density at the surface of the optically-trapped particle includes the contributions from the higher-order multipolar components, they were negligible when investigating the fluctuation of the distant probe particle. By choosing the position *r* of the investigation from the force monopole instead of the probe radius *a*, collapse of the affine elasticity similar to that observed in this study was found [35]. Although we were not aware of the mechanism behind the collapse at that time, it is now clear that the analysis procedure similar to this study will explain the collapse in the prior study, as well. Likewise, stiffening due to any order of multipolar forces at any location in the medium around the probe particle should collapse to its own single master curve. Inside of living cells, motor proteins, such as myosin and kinesins, work on the cytoskeletal networks mostly as force dipoles because the force balance is required in the typically overdamped situations. Cells possibly manifest their complex mechanical properties by integrating the nonlinear effect of intracellular force generations. The universal collapse of the affine elasticity found in this study may then help to investigate the nonlinear problem by addressing the interaction between multipolar forces in the medium, although that would go far beyond the scope of this study.

**Conclusion**

In this study, we conducted high-bandwidth 1-particle MR to investigate the microscopic mechanical response of semiflexible biopolymer networks to a localized force applied by feedback-controlled optical trapping of an embedded probe particle. By taking advantage of this high-bandwidth technology, the affine elasticity $G_A$ was obtained in a manner that is not affected by low-frequency relaxations, that should more or less occur in any hydrogel. In contrast to macrorheometry, the strains and stresses created around the externally-applied localized force is not uniform. Theoretical investigation of such heterogeneous mechanical response is challenging especially when the mechanical properties of semiflexible polymer network is highly nonlinear. However, it was found in our MR experiments that there exists a general response relationship between the applied force ($F$) and the elasticity of the surrounding medium of the probe particle ($G_A$) once they are normalized with relevant microscopic parameters that characterize the networks. The collapse of experimental data onto a single master curve was explained by the affine response theory of semiflexible biopolymer networks. We predict that deviation from the master relation may occur when smaller probe particles are used. The movement of smaller probe particle induces larger higher-order strains and the energy cost for the bending deformation of semiflexible filaments should become non-negligible. The additional energy costs required for the deformation comes into effect on the probe particle's response as it was partially found in the experiment in this study.

Mechanical properties in living cells are profoundly affected by the force-generating agents, such as molecular motors that dynamically act as localized forces on cytoskeletons [9, 10, 68, 69]. Mechanical response to the localized force is commonly heterogeneous. For instance, motor-generated forces create heterogeneously distributed fluctuations in their surrounding medium [35]. These fluctuating fields are driven by innumerous different motors in living cells and could interact each other in a complex manner, which may determine large-scale mechanical properties of active systems. It is likely that the nonlinear responses in microscales are associated with the peculiar correlations observed in the motor-driven fluctuations in active cytoskeletons [10, 48, 70, 71]. Although the nonlinearity is thought to be the origin of the enormously complex mechanics of living systems, the nonlinearity has prevented further analytical investigation. This study showed that the simple physical law summarized by the master relations exist in the apparent complexity of the local nonlinear response of semiflexible biopolymer networks [9, 10, 12, 67]. We therefore believe that this finding will help simplify the complex nonlinear problem of semiflexible biopolymer networks and elucidate how the mechanics of living cells and tissues emerge, on the basis of microscopic interactions.

**Acknowledgements**

We thank to Prof. Gijsje Koenderink, Dr. Karin Jansen at AMOLF (Netherlands) and Dr. David Head at Leeds (England) for their support on fibrin preparations and helpful discussions. This work was supported by JSPS KAKENHI Grant Number JP21H01048, JP20H05536, JP20H00128.

**Appendix A.**

In the standard theoretical model of the affine network [18, 19, 72], it is assumed that semiflexible biopolymers are crosslinked with the average length $l$. The concentration of the hydrogel is given as the length density of the filaments $\rho \sim \xi^{-2}$ ($\xi$: mesh size). Under the constraint of affinity, crosslinks in the networks change their position homogeneously by following the macroscopic deformation. Then, tensile or compressive forces arise along the filaments between the crosslinks, according to the change of its length $\delta l$. By investigating how stress emerges as the sum of these forces and how it relaxes after applying macroscopic deformations, $G(\omega)$ was predicted to be inversely proportional to the microscopic mechanical response of the semiflexible biopolymer as given by Eq. (2) [18, 19]. When a tensile force is applied to the semiflexible biopolymer, its end-to-end distance $l$ is enlarged by reducing the bending fluctuations. The entropic cost for reducing these thermal fluctuations yields the restoring resilience to this stretch, which is characterized by the bending modulus $\kappa = l_p k_B T$ ($l_p$: persistence length [72, 73]). On the other hand, the resistance to filament elongation is determined by the friction caused by the motion transverse to the filament's orientation. The transverse friction coefficient per unit length of the filament with diameter $d$ is given by [18, 72]

$$\zeta_\perp = 4\pi\eta / \ln(\xi/d) \qquad (16)$$

where $\eta$ is the viscosity of the solvent. Thus, the microscopic parameters such as $\kappa = l_p k_B T$, $l$, $\rho \sim \xi^{-2}$, and $\zeta_\perp$ relate the filament properties to $G(\omega)$.